\documentclass[journal, twocolumn]{IEEEtran}
\IEEEoverridecommandlockouts

\usepackage[final]{listings}
\usepackage{acro}

\usepackage{xifthen}

\usepackage{cite}

\usepackage{nohyperref}

\usepackage{tikz}
\usepackage{pgfplots}
\usetikzlibrary{arrows, positioning, shapes, calc, spy}
\usepgfplotslibrary{fillbetween, groupplots}

\usepackage[cmex10]{amsmath}
\usepackage{amssymb,amsfonts,amsthm}
\usepackage{array}

\usepackage{bm}

\usepackage{siunitx}
\usepackage{balance}

\usepackage{booktabs}
\usepackage{tabularx}

\usepackage{xcolor}

\usepackage{algorithm,algorithmic}

\usepackage{ifthen}

\usepackage{marginnote}

\newcommand{\internalLink}[1]{\hyperref[#1]{\ref*{#1}}}
\newcommand{\internalLinkChapter}[1]{Chapter \hyperref[#1]{\ref*{#1}}}
\newcommand{\internalLinkSection}[1]{Section \hyperref[#1]{\ref*{#1}}}
\newcommand{\internalLinkAppendix}[1]{Appendix \hyperref[#1]{\ref*{#1}}}
\newcommand{\internalTab}[1]{Table~\hyperref[#1]{\ref*{#1}}}
\newcommand{\internalFig}[1]{Fig.~\hyperref[#1]{\ref*{#1}}}
\newcommand{\internalEq}[1]{(\hyperref[#1]{\ref*{#1}})}
\newcommand{\internalEqs}[2]{(\hyperref[#1]{\ref*{#1}})-(\hyperref[#2]{\ref*{#2}})}
\newcommand{\internalListing}[1]{Listing \hyperref[#1]{\ref*{#1}}}
\newcommand{\internalAlgorithm}[1]{Algorithm~\hyperref[#1]{\ref*{#1}}}
\newcommand{\internalLemma}[1]{Lemma \hyperref[#1]{\ref*{#1}}}
\newcommand{\internalDefinition}[1]{Def. \hyperref[#1]{\ref*{#1}}}
\newcommand{\internalTheorem}[1]{Theorem \hyperref[#1]{\ref*{#1}}}

\let\originalleft\left
\let\originalright\right
\renewcommand{\left}{\mathopen{}\mathclose\bgroup\originalleft}
\renewcommand{\right}{\aftergroup\egroup\originalright}

\newcommand{\sign}[1]{\operatorname{sign}\left(#1\right)}
\newcommand{\relu}[1]{\operatorname{ReLU}\left(#1\right)}

\newcommand{\RM}{\operatorname{RM}}

\newcommand{\EbNoLine}{E_\mathsf{b}/N_0}

\DeclareSIUnit{\belmilliwatt}{Bm}
\DeclareSIUnit{\dBm}{\deci\belmilliwatt}
\DeclareSIUnit{\bit}{bit}
\DeclareSIUnit{\bits}{bits}

\newcolumntype{C}{>{{}}c<{{}}}
\DeclareAcronym{BCH}{
	short = BCH,
	long= Bose-Chaudhuri-Hocquenghem
}

\DeclareAcronym{BER}{
	short = BER,
	long= bit error rate
}

\DeclareAcronym{BLER}{
	short = BLER,
	long= block error rate
}

\DeclareAcronym{BP}{
	short = BP,
	long = belief propagation
}

\DeclareAcronym{CN}{
	short = CN,
	long = check node
}

\DeclareAcronym{LDPC}{
	short = LDPC,
	long = low-density parity-check
}

\DeclareAcronym{LLR}{
	short = LLR,
	long = log-likelihood ratio
}

\DeclareAcronym{MBBP}{
	short = MBBP,
	long = multiple-bases belief propagation
}

\DeclareAcronym{NBP}{
	short = NBP,
	long = neural belief propagation
}

\DeclareAcronym{NOMS}{
	short = NOMS,
	long = neural offset min-sum
}

\DeclareAcronym{ML}{
	short = ML,
	long = maximum-likelihood
}

\DeclareAcronym{relu}{
	short = reLu,
	long = rectified linear unit
}

\DeclareAcronym{RM}{
	short = RM,
	long= Reed-Muller
}

\DeclareAcronym{RNN}{
	short = RNN,
	long= recurrent neural network
}

\DeclareAcronym{STE}{
	short = STE,
	long = straight-through estimator
}

\DeclareAcronym{PBNBP}{
	short = PB-NBP,
	long = pruning-based neural belief propagation
}

\DeclareAcronym{PBNOMS}{
	short = PB-NOMS,
	long = pruning-based neural offset min-sum
}

\DeclareAcronym{VN}{
	short = VN,
	long = variable node
}

\newcommand{\tablescaling}{1.0}
\ifCLASSOPTIONonecolumn
  \renewcommand{\tablescaling}{0.5}
\fi
\newcommand{\twocoleqbreak}[1]{}
\ifCLASSOPTIONtwocolumn
  \renewcommand{\twocoleqbreak}[1]{\nonumber\\ #1}
\fi

\begin{document}
\title{Pruning and Quantizing\\Neural Belief Propagation Decoders}

\author{\begin{center}Andreas~Buchberger,~\IEEEmembership{Student~Member,~IEEE,} Christian Häger,~\IEEEmembership{Member,~IEEE,}\\Henry D. Pfister,~\IEEEmembership{Senior~Member,~IEEE,}  Laurent~Schmalen,~\IEEEmembership{Senior~Member,~IEEE,} and~Alexandre~Graell~i~Amat,~\IEEEmembership{Senior~Member,~IEEE}\end{center}
\thanks{This work was presented in part at the \emph{IEEE International Symposium on Information  Theory (ISIT) 2020.}}
\thanks{This work was partially funded by the European Union's Horizon 2020 research and innovation programme under the Marie Sk\l{}odowska-Curie grant agreements no. 676448 and no. 749798 and by the Swedish Research Council under grant 2016-04253. Parts of the simulations were performed on resources at C3SE provided by the Swedish national infrastructure for computing.}
\thanks{A. Buchberger, C. Häger, and A. Graell i Amat are with the Department of Electrical Engineering, Chalmers University of Technology, Gothenburg, SE--412 96, Sweden, e-mail: \mbox{\{firstname.lastname\}@chalmers.se.}}
\thanks{H. D. Pfister is with the Department of Electrical and Computer Engineering, Duke University, Durham, North Carolina, USA, e-mail: \mbox{henry.pfister@duke.edu}}
\thanks{L. Schmalen is with the Communications Engineering Lab, \mbox{Karlsruhe} Institute of Technology (KIT), 76131 Karlsruhe, Germany, e-mail: \mbox{laurent.schmalen@kit.edu}}
}

\maketitle
\acresetall

%%%%%%%%%%%%%%%%%%%%%%%%%%%%%%%%%%%%%%%%%%%%%%%%%%%%%%%%%%%%%%%%%%%%%%%%%%%%%%%
%                                                                             %
%                                   ABSTRACT                                  %
%                                                                             %
%%%%%%%%%%%%%%%%%%%%%%%%%%%%%%%%%%%%%%%%%%%%%%%%%%%%%%%%%%%%%%%%%%%%%%%%%%%%%%%

\begin{abstract}
We consider near \ac{ML} decoding of short linear block codes. In particular, we propose a novel decoding approach based on \ac{NBP} decoding recently introduced by Nachmani \emph{et al.} in which we allow a different parity-check matrix in each iteration of the algorithm.  The key idea is to consider  \ac{NBP} decoding over an overcomplete parity-check matrix and use  the weights of \ac{NBP} as a measure of the importance of the  \acp{CN} to decoding. The unimportant \acp{CN} are then pruned. In contrast to NBP, which performs decoding on a given \emph{fixed} parity-check matrix, the proposed \ac{PBNBP} typically results in a different parity-check matrix in each iteration. For a given complexity in terms of CN evaluations, we show that \ac{PBNBP} yields significant performance improvements with respect to NBP.
We apply the proposed decoder to the decoding of  a \acl{RM} code, a short \ac{LDPC} code, and a polar code.  \ac{PBNBP}   outperforms  \ac{NBP} decoding over an overcomplete parity-check matrix by 0.27--0.31~dB while reducing the number of required \ac{CN} evaluations by up to 97\%. For the \ac{LDPC} code,  \ac{PBNBP} outperforms conventional \acl{BP} with the same number of \ac{CN} evaluations by 0.52~dB.
We further extend the pruning concept to offset min-sum decoding and introduce a \ac{PBNOMS} decoder, for which we jointly optimize the offsets and the quantization of the messages and offsets. We demonstrate performance 0.5~dB from \ac{ML} decoding with 5-bit quantization for the \acl{RM} code.
\end{abstract}
\acresetall

%%%%%%%%%%%%%%%%%%%%%%%%%%%%%%%%%%%%%%%%%%%%%%%%%%%%%%%%%%%%%%%%%%%%%%%%%%%%%%%
%                                                                             %
%                                 INTRODUCTION                                %
%                                                                             %
%%%%%%%%%%%%%%%%%%%%%%%%%%%%%%%%%%%%%%%%%%%%%%%%%%%%%%%%%%%%%%%%%%%%%%%%%%%%%%%
\vskip 10pt
\section{Introduction}
\IEEEPARstart{F}{or} short code lengths, algebraic codes such as \ac{BCH} codes and \ac{RM} codes show excellent performance under \ac{ML} decoding. However, achieving near-\ac{ML} performance using conventional methods is computationally complex.
Fueled by the advances in the field of deep learning, deep neural networks have also gained interest in the coding community \cite{Nachmani2016,Nachmani2018, Lian2019,Gruber2017,Lugosch2017}. In \cite{Nachmani2018}, \ac{BP} decoding is formulated in the context of deep neural networks. Instead of iterating between \acp{CN} and \acp{VN}, the messages are passed through unrolled iterations in a feed-forward fashion. Additionally, weights can be introduced at the edges, which are then optimized using stochastic gradient descent (and variants thereof). This decoding method is commonly referred to as \ac{NBP} and can be seen as a generalization of \ac{BP} decoding where all individual messages are scaled by a single damping coefficient \cite{Halford2006}. The weights in \ac{NBP} can counteract the effect of short cycles by scaling messages accordingly. The concept of \ac{NBP} is extended to \ac{NOMS} in \cite{Lugosch2017} by assigning an individual offset to each edge of the unrolled graph for the \ac{CN} update \cite{Fossorier1999,Chen2002}.

While \ac{NBP} and \ac{NOMS} decoding improve upon conventional \ac{BP} and offset min-sum decoding, their performance is still limited by the underlying parity-check matrix. Different parity-check matrices may yield different performances. This fact has been exploited by using redundant parity-check matrices \cite{Bossert1986,Kothiyal2005,Jiang2006,Halford2006,Hehn2010,Santi2018}.  In particular, \cite{Hehn2010} proposed \ac{MBBP}, which selects the best decoded codeword from multiple parallel \ac{BP} decoders over different parity-check matrices. For the decoding of \ac{RM} codes, \cite{Santi2018} considered applying \ac{BP} with a single damping coefficient (a single weight) to the parity-check matrix composed of all minimum-weight parity checks. In \cite{Lian2019}, \ac{NBP} over parity-check matrices containing all minimum-weight parity checks was investigated. While using large, redundant parity-check matrices yields excellent performance close to \ac{ML}, it suffers from high computational complexity.

In this paper, we propose \ac{PBNBP}, a novel decoding approach based on NBP decoding to selecting the best parity-check equations for each iteration of the algorithm.
The proposed approach starts with \ac{NBP} decoding over the unrolled graph corresponding to a large overcomplete parity-check matrix of the linear block code. The key idea is to interpret the trained weights as a measure of the contribution of the corresponding \acp{CN} to the decoding process. CNs with small contribution to the decoding are then pruned. More precisely, we tie the weights of all edges emanating from a  \ac{CN}. \acp{CN} connected to low-magnitude-weight edges do not play an important role in the decoding process and are pruned.
 Pruning results in an unrolled graph with a different set of \acp{CN} in each \ac{CN} layer. This corresponds to using a different parity-check matrix for each iteration of \ac{NBP}.   We investigate three variants of the \ac{PBNBP} decoder\textemdash untying all weights in the resulting (unrolled) Tanner graph, using the weights obtained during the optimization process directly, and setting all weights to one. For Reed-Muller codes,  we show that PB-BP decoding outperforms NBP over the overcomplete matrix and \ac{MBBP} and achieves near-ML performance. Moreover, the pruning results in a lower-complexity decoder compared to NBP over the overcomplete matrix.  We also give results for an LDPC code and a polar code.

We further extend the pruning concept to \ac{NOMS} decoding, leading to the formulation of  \ac{PBNOMS} decoding. For \ac{PBNOMS} we  investigate the joint quantization of the weights, offsets, channel messages, and messages between layers.   We use a \ac{STE} \cite{Bengio2013,Yin2019} to define the gradient of the quantizer and let the quantization levels and thresholds be trainable. For various codes, we illustrate the  performance of \ac{PBNOMS} for different quantizations.

%%%%%%%%%%%%%%%%%%%%%%%%%%%%%%%%%%%%%%%%%%%%%%%%%%%%%%%%%%%%%%%%%%%%%%%%%%%%%%%
%                                                                             %
%                                PRELIMINARIES                                %
%                                                                             %
%%%%%%%%%%%%%%%%%%%%%%%%%%%%%%%%%%%%%%%%%%%%%%%%%%%%%%%%%%%%%%%%%%%%%%%%%%%%%%%
\section{Preliminaries}
\label{sec:preliminaries}
Consider a linear block code \(\mathcal{C}\) of length \(n\) and dimension \(k\) with parity-check matrix \(\bm{H}\) of size \(m \times n\), \(m \ge n - k\). If \(m > n - k\), we refer to the parity-check matrix as an overcomplete matrix and denote it as \(\bm{H}_\mathsf{oc}\). The case \(m=n-k\) corresponds to a parity-check matrix with no redundant rows, which we refer to as \(\bm{H}_\mathsf{std}\). We denote the Tanner graph corresponding to a parity-check matrix as \(\mathcal{G} = (\mathcal{V}_\mathsf{v}, \mathcal{V}_\mathsf{c}, \mathcal{E})\), consisting of a set of \(m\) \acp{CN}, \(\mathcal{V}_\mathsf{c}=\{\mathsf{c}_1,\ldots,\mathsf{c}_m\}\), a set of \(n\) \acp{VN}, \(\mathcal{V}_\mathsf{v}=\{\mathsf{v}_1,\ldots,\mathsf{v}_n\}\), and a set of edges \(\mathcal{E}\) connecting \acp{CN} with \acp{VN}.

For each \ac{VN} \(\mathsf{v}\in \mathcal{V}_\mathsf{v}\) we define its neighborhood
\begin{align*}
  \mathcal{N}(\mathsf{v}) &\triangleq \left\{\mathsf{c}\in\mathcal{V}_\mathsf{c}:(\mathsf{v},\mathsf{c})\in \mathcal{E}\right\}
\end{align*}
i.e., the set of all \acp{CN} connected to \ac{VN} \(\mathsf{v}\). Equivalently, we define the neighborhood of a \ac{CN} \(\mathsf{c}\in \mathcal{V}_\mathsf{c}\) as
\begin{align*}
  \mathcal{N}(\mathsf{c}) &\triangleq \left\{\mathsf{v}\in\mathcal{V}_\mathsf{v}:(\mathsf{v},\mathsf{c})\in \mathcal{E}\right\}.
\end{align*}
Let \(\mu_{\mathsf{v}_i\rightarrow\mathsf{c}_j}^{(\ell)}\)  be the message passed from \ac{VN} \(\mathsf{v}_i\in\mathcal{V}_\mathsf{v}\) to \ac{CN} \(\mathsf{c}_j\in\mathcal{V}_\mathsf{c}\) and \(\mu_{\mathsf{c}_j\rightarrow\mathsf{v}_i}^{(\ell)}\) the message passed from \ac{CN} \(\mathsf{c}_j\in\mathcal{V}_\mathsf{c}\) to \ac{VN} \(\mathsf{v}_i\in\mathcal{V}_\mathsf{v}\) in the \(\ell\)-th decoding iteration. For \ac{BP} decoding, the \ac{VN} and \ac{CN} updates are
\begin{align}
  \mu_{\mathsf{v}_i\rightarrow\mathsf{c}_j}^{(\ell)} &=  \mu_{\mathsf{ch},\mathsf{v}_i} + \sum_{\mathsf{c}\in \mathcal{N}(\mathsf{v}_i)\backslash \mathsf{c}_j}  \mu_{\mathsf{c}\rightarrow\mathsf{v}_i}^{(\ell)}
  \label{eq:vn_update}
\end{align}
and
\begin{align}
  \mu_{\mathsf{c}_j\rightarrow\mathsf{v}_i}^{(\ell)} &= 2\tanh^{-1}\left(\prod_{\mathsf{v}\in \mathcal{N}(\mathsf{c}_j)\backslash \mathsf{v}_i}  \tanh\left(\frac{1}{2}\mu_{\mathsf{v}\rightarrow\mathsf{c}_j}^{(\ell)}\right)\right)
  \label{eq:cn_update}
\end{align}
respectively, where \(\mu_{\mathsf{ch},\mathsf{v}_i}\) is the channel message. For binary transmission over the additive white Gaussian noise channel
\begin{align*}
 \mu_{\mathsf{ch},\mathsf{v}_i} &\triangleq \ln \frac{p_{Y|B}(y_i|b_i=0)}{p_{Y|B}(y_i|b_i=1)} \overset{}{=}  \frac{2y_i}{\sigma^2}
\end{align*}
where \(y_i\) is the channel output, \(b_i\) is the transmitted bit, and \(\sigma^2\) is the noise variance.
The \emph{a posteriori} \ac{LLR} in the \(\ell\)-th iteration is
\begin{align*}
  \mu_{\mathsf{v}_i}^{(\ell)} &=  \mu_{\mathsf{ch},\mathsf{v}_i} + \sum_{\mathsf{c}\in \mathcal{N}(\mathsf{v}_i)}  \mu_{\mathsf{c}\rightarrow\mathsf{v}_i}^{(\ell)}.
  % \label{eq:vn_marginalization}
\end{align*}

A large contribution to the computational complexity stems from the \ac{CN} update \internalEq{eq:cn_update}. A widely-used low-complexity approximation to \internalEq{eq:cn_update} is the min-sum approximation \cite{Fossorier1999}
\begin{align*}
  \mu_{\mathsf{c}_j\rightarrow\mathsf{v}_i}^{(\ell)} &= \min_{\mathsf{v}\in \mathcal{N}(\mathsf{c}_j)\backslash \mathsf{v}_i}\left|\mu_{\mathsf{v}\rightarrow \mathsf{c}_j}^{(\ell)}\right| \prod_{\mathsf{v}\in \mathcal{N}(\mathsf{c}_j)\backslash \mathsf{v}_i}  \sign{\mu_{\mathsf{v}\rightarrow \mathsf{c}_j}^{(\ell)}}
  % \label{eq:cn_min_sum}
\end{align*}
where \(\sign{\cdot}\) denotes the sign function.
As this approximation tends to overestimate the magnitude of the messages, an additive \ac{CN}- and iteration-dependent correction factor \(\beta_{\mathsf{c}_j}^{(\ell)}\) is often introduced, leading to offset min-sum decoding \cite{Chen2002}
\begin{align}
  \mu_{\mathsf{c}_j\rightarrow\mathsf{v}_i}^{(\ell)} &= \max\left(\hspace{-2pt}\min_{\mathsf{v}\in \mathcal{N}(\mathsf{c}_j)\backslash \mathsf{v}_i}\left|\mu_{\mathsf{v}\rightarrow \mathsf{c}_j}^{(\ell)}\right|-\beta_{\mathsf{c}_j}^{(\ell)}, 0\hspace{-2pt}\right) \twocoleqbreak{&\quad\qquad\qquad\qquad\qquad \cdot\hspace{-10pt}}
  \prod_{\mathsf{v}\in \mathcal{N}(\mathsf{c}_j)\backslash \mathsf{v}_i}  \sign{\mu_{\mathsf{v}\rightarrow \mathsf{c}_j}^{(\ell)}}.
  \label{eq:cn_offset_min_sum}
\end{align}

\subsection{Neural Belief Propagation}
\begin{figure}
  \centering
  \includegraphics{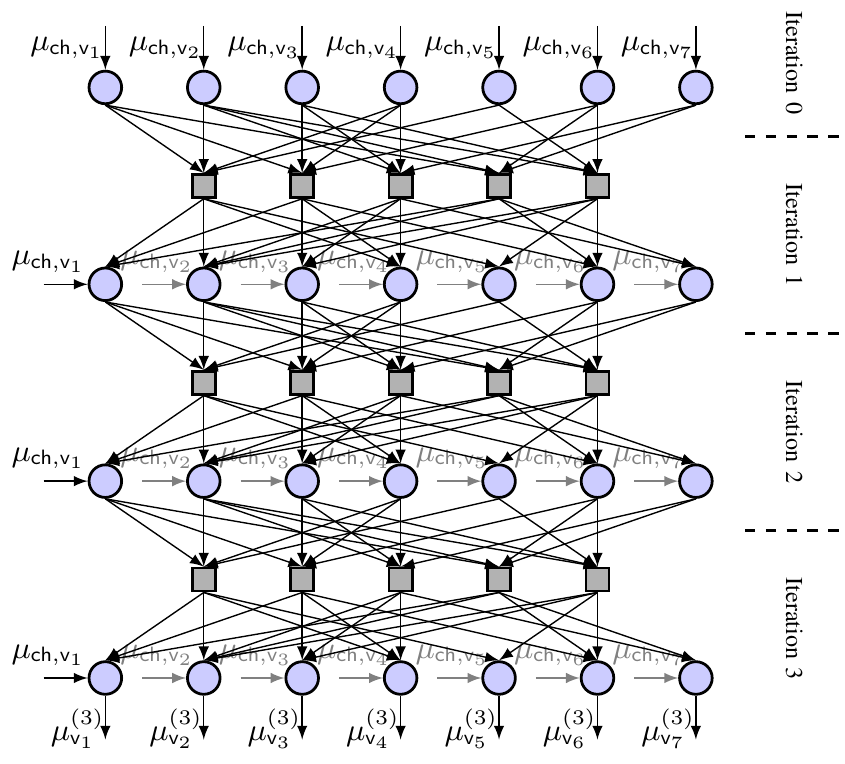}
  \caption{Unrolled graph for three iterations where the blue circles denote \acp{VN} and the gray squares \acp{CN}.}
  \label{fig:unrolled_graph}
\end{figure}

For conventional \ac{BP}, the decoder iterates between \ac{VN} and \ac{CN} updates by passing messages along the edges connecting \acp{VN} and \acp{CN}. For a given number of iterations \(\ell_\mathsf{max}\), one can \emph{unroll} the graph by stacking \(\ell_\mathsf{max}\) copies of the Tanner graph. Consequently, the messages are passed in an unrolled fashion through the graph. This is illustrated in \internalFig{fig:unrolled_graph} for three iterations. One way to counteract the effect of short cycles on the performance of \ac{BP} decoding for short linear block codes is to introduce weights for each edge of the unrolled Tanner graph \cite{Nachmani2016,Nachmani2018}. Due to the similarities of the weighted unrolled graph with a neural network, this is referred to as  \ac{NBP}. For \ac{NBP}, the update rules \internalEq{eq:vn_update} and \internalEq{eq:cn_update} are modified to
\begin{align}
  \mu_{\mathsf{v}_i\rightarrow\mathsf{c}_j}^{(\ell)} &=w_{\mathsf{v}_i\rightarrow\mathsf{c}_j}^{(\ell)}\left( w_{\mathsf{ch},\mathsf{v}_i}^{(\ell)}\mu_{\mathsf{ch},\mathsf{v}_i} + \sum_{\mathsf{c}\in \mathcal{N}(\mathsf{v}_i)\backslash \mathsf{c}_j}  \mu_{\mathsf{c}\rightarrow\mathsf{v}_i}^{(\ell)}\right)
  \label{eq:vn_update_wbp}
\end{align}
and
\begin{align}
\mu_{\mathsf{c}_j\rightarrow\mathsf{v}_i}^{(\ell)} &= 2 w_{\mathsf{c}_j\rightarrow\mathsf{v}_i}^{(\ell)}\tanh^{-1}\left(\prod_{\mathsf{v}\in \mathcal{N}(\mathsf{c}_j)\backslash \mathsf{v}_i} \hspace{-12pt}\tanh\left(\frac{1}{2}\mu_{\mathsf{v}\rightarrow\mathsf{c}_j}^{(\ell)}\right)\hspace{-3pt}\right)
  \label{eq:cn_update_wbp}
\end{align}
where \( w_{\mathsf{ch},\mathsf{v}}^{(\ell)}\), \(w_{\mathsf{v}\rightarrow\mathsf{c}}^{(\ell)}\), and \(w_{\mathsf{c}\rightarrow\mathsf{v}}^{(\ell)}\), are the channel weights, the weights on the edges connecting \acp{VN} to \acp{CN}, and the weights on the edges connecting \acp{CN} to \acp{VN}, respectively.
The \emph{a posteriori} \ac{LLR} in the \(\ell\)-th iteration is
\begin{align*}
  \mu_{\mathsf{v}_i}^{(\ell)} &= w_{\mathsf{ch},\mathsf{v}_i}^{(\ell)}\mu_{\mathsf{ch},\mathsf{v}_i} + \sum_{\mathsf{c}\in \mathcal{N}(\mathsf{v}_i)}  \mu_{\mathsf{c}\rightarrow\mathsf{v}_i}^{(\ell)}.
  % \label{eq:vn_marginlaization_wbp}
\end{align*}
In \internalEq{eq:vn_update_wbp} and \internalEq{eq:cn_update_wbp} the weights are untied over all nodes as well as over all iterations, i.e., each edge has an individual weight. In order to reduce complexity and storage requirements for \ac{NBP}, the weights can also be tied. In \cite{Lian2019}, tying the weights temporally, i.e., over  iterations, and spatially, i.e., all edges within a layer have the same weight, was explored.
Note that setting all  weights of an \ac{NBP} decoder to one yields conventional \ac{BP} decoding.

\subsection{Neural Offset Min-Sum Decoder}
Similar to the extension of conventional \ac{BP} decoding to \ac{NBP} decoding, the offset min-sum decoder can be extended to \iac{NOMS} decoder \cite{Lugosch2017}. Instead of the \ac{CN}- and iteration-dependent offset \(\beta_\mathsf{c}^{(\ell)}\), each edge emanating from a \ac{CN} has its own offset,
\begin{align}
  \mu_{\mathsf{c}_j\rightarrow\mathsf{v}_i}^{(\ell)}&= \relu{\min_{\mathsf{v}\in \mathcal{N}(\mathsf{c}_j)\backslash \mathsf{v}_i}\left|\mu_{\mathsf{v}\rightarrow \mathsf{c}_j}^{(\ell)}\right|-\beta_{\mathsf{c}_j\rightarrow\mathsf{v}_i}^{(\ell)}}\twocoleqbreak{ & \quad\qquad\qquad\qquad\qquad\cdot\hspace{-10pt}}
  \prod_{\mathsf{v}\in \mathcal{N}(\mathsf{c}_j)\backslash \mathsf{v}_i}  \sign{\mu_{\mathsf{v}\rightarrow \mathsf{c}_j}^{(\ell)}}
  \label{eq:cn_update_noms}
\end{align}
where \(\relu{\cdot} = \max(\cdot, 0)\) denotes a rectified activation function commonly used in neural networks \cite{Nair2010}.

%%%%%%%%%%%%%%%%%%%%%%%%%%%%%%%%%%%%%%%%%%%%%%%%%%%%%%%%%%%%%%%%%%%%%%%%%%%%%%%
%                                                                             %
%                                    PRUNING                                  %
%                                                                             %
%%%%%%%%%%%%%%%%%%%%%%%%%%%%%%%%%%%%%%%%%%%%%%%%%%%%%%%%%%%%%%%%%%%%%%%%%%%%%%%
\section{ Pruning-Based Neural Belief Propagation Decoders}
\label{sec:pruning}
Here we propose \ac{PBNBP}. The main idea is to consider \ac{NBP} over the unrolled graph starting from a large, overcomplete parity-check matrix \(\bm{H}_\mathsf{oc}\). We tie the weights for each \ac{CN}, i.e., \(w_{\mathsf{c}_j\rightarrow\mathsf{v}_i}^{(\ell)} = w_{\mathsf{c}_j}^{(\ell)}\) for all \(\mathsf{v}_i\in\mathcal{N}(\mathsf{c}_j)\),
\begin{align}
  \mu_{\mathsf{c}_j\rightarrow\mathsf{v}_i}^{(\ell)} &= 2 w_{\mathsf{c}_j}^{(\ell)}\tanh^{-1}\left(\prod_{\mathsf{v}\in \mathcal{N}(\mathsf{c}_j)\backslash \mathsf{v}_i} \hspace{-12pt}\tanh\left(\frac{1}{2}\mu_{\mathsf{v}\rightarrow\mathsf{c}_j}^{(\ell)}\right)\hspace{-2pt}\right).
  \label{eq:cn_update_pp}
\end{align}
We view the weights as an indication of the importance of the respective \ac{CN} to the decoding and use them to prune the graph by successively removing \acp{CN} associated to low weights. The resulting graph potentially consists of a different set of \acp{CN} at each \ac{CN} layer. This corresponds to  selecting a (potentially) different set of parity-check equations from the overcomplete parity-check matrix in each iteration of \ac{BP} decoding.

We first describe how we optimize the graph's weights and in a second step, we present the training procedure to prune the graph.

\subsection{Optimization of the Weights}
The decoding process can be seen as a classification task where the channel output is mapped to a valid codeword. This task consists of \(2^k\) classes, one for each codeword. Training such a classification task is infeasible as the resulting decoder typically generalizes poorly to classes not contained in the training data \cite{Gruber2017}. Alternatively, one may consider a binary classification task for each of the \(n\) bits. As a loss function, the average bitwise cross-entropy between the transmitted bits and the \ac{VN} output \acp{LLR} of the final \ac{VN} layer can be used \cite{Nachmani2016,Nachmani2018},
\begin{align*}
\Gamma &=  -\frac{1}{n}\sum_{i=1}^n \log\left(o_i^{b_i} \left(1-o_i\right)^{1-b_i}\right)
% \label{eq:loss_x_entropy}
\end{align*}
where \(o_i\) is the estimate of the probability that the \(i\)-th bit after the final iteration is one,
\begin{align*}
  o_i &=  \frac{e^{-\mu_{\mathsf{v}_i}^{(\ell_\mathsf{max})}}}{1+e^{-\mu_{\mathsf{v}_i}^{(\ell_\mathsf{max})}}},
\end{align*}
and \(\ell_\mathsf{max}\) denotes the number of decoding iterations.
The optimization behavior can be improved by using a multiloss function \cite{Nachmani2016,Nachmani2018}, where the overall loss is the average  bitwise cross-entropy between the transmitted bits and the \ac{VN} output \ac{LLR} of each \ac{VN} layer.
The cross-entropy is well-suited for the bitwise classification task, but it does not necessarily result in a decoder with the lowest possible bit error rate. In fact, the bit error rate would be a more natural choice for the loss function. However, since the gradients would be zero almost everywhere, it is infeasible for optimization using gradient descent. Instead, \cite{Lian2019} proposed  a new loss function which can be regarded as \emph{soft  bit error rate},
\begin{align}
\Gamma &=  \frac{1}{n}\sum_{i=1}^n \left(1-o_i\right)^{b_i} o_i^{1-b_i}.
\label{eq:loss_soft_ber}
\end{align}
It was empirically observed in \cite{Lian2019} that minimizing this loss function also minimizes the bit error rate.
Combining the soft bit error rate and a multiloss results in
\begin{align}
\tilde{\Gamma} &= \frac{1}{\sum_\ell \eta^{{\ell_\mathsf{max}}
-\ell}}\sum_{\ell=1}^{\ell_\mathsf{max}}
 \eta^{{\ell_\mathsf{max}}
-\ell} \frac{1}{n}\sum_{i=1}^n \left(1-o_i^{(\ell)}\right)^{b_i} \left(o_i^{(\ell)}\right)^{1-b_i}\nonumber\\
&\overset{(a)}{=}\frac{1}{\sum_\ell \eta^{{\ell_\mathsf{max}}
-\ell}}\sum_{\ell=1}^{\ell_\mathsf{max}}
 \eta^{{\ell_\mathsf{max}}
-\ell} \frac{1}{n}\sum_{i=1}^n  o_i^{(\ell)}
\label{eq:loss_soft_ber_mult}
\end{align}
where \(o_i^{(\ell)}\) is the estimate of the probability that the \(i\)-th bit is one after the \(\ell\)-th iteration, and \(\eta\in(0;1]\) determines the contribution of intermediate layers to the overall loss. Step \((a)\) follows from the assumption that the all-zero codeword is transmitted, which is a valid assumption since the channel and the decoder are symmetric.
The parameter \(\eta\)  is set close to one, i.e., all layers contribute equally to the loss, in the beginning of the training. This allows for gradients to efficiently propagate to earlier layers and hence improves convergence. However, this does not correspond to the desired, final behavior of the decoder where only the output of the final layer matters. Thus, the contribution of the intermediate layers is successively decreased during training by means of decreasing \(\eta\). During the final stages of training, only the last layer will contribute to the loss, corresponding to the desired, final behavior \cite{Lian2019}.
\begin{figure}[t]
    \centering
    \includegraphics{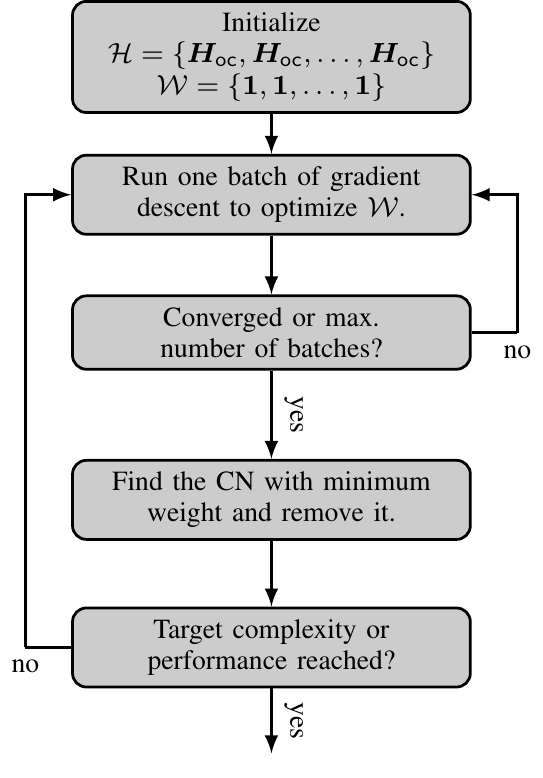}
    \caption{Flowchart of the training process. \(\bm{1}\) denotes the all-one matrix of appropriate size.}
    \label{fig:training}
\end{figure}
\subsection{Training Procedure}
\label{sec:training}
Consider \ac{NBP} with \ac{VN} update \internalEq{eq:vn_update_wbp} and  \ac{CN} update \internalEq{eq:cn_update_pp} over the unrolled graph of \(\bm{H}_\mathsf{oc}\)  in which the weights are tied at the \acp{CN}, i.e., all messages emanating from a single \ac{CN} \(\mathsf{c}\) are weighted by the same weight \(w_\mathsf{c}^{(\ell)}\). The magnitude of the weight \(w_\mathsf{c}^{(\ell)}\) can now be interpreted as a measure of how much \ac{CN} \(\mathsf{c}\) contributes to the decoding at iteration \(\ell\).  A large magnitude indicates high importance whereas a magnitude of zero indicates that the \ac{CN} is irrelevant to the decoding process.

Let \(\bm{H}_\ell\) be the parity-check matrix used for decoding in the \(\ell\)-th iteration and define \(\mathcal{H} = \{\bm{H}_1, \ldots, \bm{H}_{\ell_\mathsf{max}}\}\). The set is initialized with the same large overcomplete matrix \(\bm{H}_\mathsf{oc}\) for each iteration, i.e., \(\mathcal{H} = \{\bm{H}_\mathsf{oc}, \ldots, \bm{H}_\mathsf{oc}\}\).
Equivalently, we define a set of weights \(\mathcal{W}\) and initialize all weights to one, i.e., we start with conventional \ac{BP}. The weights in \(\mathcal{W}\) are then optimized using the Adam optimizer \cite{Kingma2014} within the Tensorflow programming framework \cite{tensorflow2015}.  After the optimization has converged, we find the lowest \ac{CN} weight \(w_\mathsf{c}^{(\ell)}\) and set it to zero. This is equivalent to pruning the \ac{CN} from the unrolled graph. As this may change the optimal value for the remaining weights, we rerun the training.
We iterate between retraining and pruning \acp{CN} and monitor the loss. The optimization is stopped when the loss starts increasing. Empirically we observe that the loss is not monotone and that it is beneficial to allow the loss to increase slightly before stopping the pruning. Alternatively, one may further prune \acp{CN} to reach a desired number of parity-check equations over all iterations, i.e., a given complexity, at the expense of a larger loss.
The result of the optimization is a set of parity-check matrices \(\mathcal{H}_\mathsf{opt} = \left\{\bm{H}_{1,\mathsf{opt}}, \ldots, \bm{H}_{\ell_\mathsf{max}, \mathsf{opt}}\right\}\) and optimized weights \(\mathcal{W}_\mathsf{opt}\). The training process is  illustrated in the flowchart of  \internalFig{fig:training} and in Algorithm 1.

\begin{algorithm}[t]
 \caption{Training process.}
 \label{lst:alg_training}
 \begin{algorithmic}[1]
   \renewcommand\algorithmicdo{\textbf{:}}
   \renewcommand\algorithmicthen{\textbf{:}}
   \renewcommand\algorithmicelse{\textbf{else :}}
  \STATE \(\mathcal{H} = \{\bm{H}_\mathsf{oc}, \ldots, \bm{H}_\mathsf{oc}\}\)
  \STATE \(\mathcal{W} = \{\bm{1}, \ldots, \bm{1}\}\)
  \STATE {\emph{Choose abort-criterion:} Prune a certain number of \acp{CN} \textbf{or} stop when the loss starts to increase}
  \WHILE {abort-criterion not fulfilled}
    \FOR {\# of batches}
      \STATE Sample \(\bm{\mu}_\mathsf{ch}\) according to channel model
      \STATE Decode \(\bm{\mu}_\mathsf{ch}\) with decoder (\(\mathcal{H}, \mathcal{W}\))
      \STATE Calculate loss \(\tilde{\Gamma}\) using \internalEq{eq:loss_soft_ber_mult}
      \STATE Calculate gradients \(\partial\tilde{\Gamma} / \partial w\) for all \(w\in\mathcal{W}\)
      \STATE Update \(\mathcal{W}\) using Adam
      \IF {Avg. loss over \(100\) batches has not improved}
      \STATE {Break}
      \ENDIF
    \ENDFOR
    \STATE {Find the smallest CN weight and remove the corresponding parity-check equation from the corresponding parity-check matrix in  \(\mathcal{H}\).}
  \ENDWHILE
  \RETURN \(\mathcal{H}\), \(\mathcal{W}\)
 \end{algorithmic}
 \end{algorithm}

If the matrices used to initialize \(\mathcal{H}\) are very large, this way of optimizing the parity-check matrices is potentially complex and slow. Empirically, we observed that it is possible to simultaneously prune more than one \ac{CN} in the earlier stages of the pruning process without harming the final performance. This allows for a significant speed-up of the optimization. All results in this paper are achieved by only pruning a single \ac{CN} per pruning step. An in-depth analysis of different pruning schedules is left for future work.

 \subsection{Complexity Discussion}
  In the following, we provide a high-level discussion of the decoding  complexity. A thorough complexity analysis would require considering hardware implementation details such as data bus requirements, impact of the degree of parallelism and structure in the graph, etc. While this is out of the scope of the paper, we note that hardware constraints can  potentially be incorporated in the training process (e.g., through a modified loss function), rendering our proposed decoders adaptable to different use cases.

  On a high level, the \ac{CN} update is the most complex operation in a \ac{BP} decoder due to the evaluation of the  \(\tanh\) and inverse \(\tanh\) functions. Hence, a commonly used measure for complexity is given by \cite{Smith2010}
   \begin{align}
     \sum_\ell  \overline{d}_\mathsf{c}^{(\ell)}\left|\mathcal{V}_{\mathsf{c}}^{(\ell)}\right|
     \label{eq:complexity}
   \end{align}
   where \(\overline{d}_\mathsf{c}^{(\ell)}\) denotes the average \ac{CN} degree in the \(\ell\)-th iteration and \(\mathcal{V}_\mathsf{c}^{(\ell)}\) the set of active \acp{CN} in the \(\ell\)-th iteration. For conventional \ac{BP} decoding, \(\mathcal{V}_{\mathsf{c}}^{(\ell)} = \mathcal{V}_{\mathsf{c}}\) and \(\overline{d}_\mathsf{c}^{(\ell)} = \overline{d}_\mathsf{c}~\forall \ell\).

   The required memory is related to the parity-check matrix itself and the number of weights that need to be stored. Since the weights are real numbers as opposed to binary values for the edges, we quantify memory requirements with the number of weights. Furthermore, the weights require additional multiplications.

   Following this discussion, we define three decoders of different complexity.
   \begin{itemize}
     \item \ac{PBNBP} decoder \(\mathcal{D}_1\): It uses the result from the optimization directly, i.e.,   \(\mathcal{H}_\mathsf{opt}\) and \(\mathcal{W}_\mathsf{opt}\).
     \item \ac{PBNBP} decoder \(\mathcal{D}_2\): It uses the optimized set of parity-check matrices, i.e.,  \(\mathcal{H}_\mathsf{opt}\), but sets all weights to one, i.e., neglects  \(\mathcal{W}_\mathsf{opt}\).
     \item \ac{PBNBP} decoder \(\mathcal{D}_3\):  It uses the optimized set of parity-check matrices, i.e.,  \(\mathcal{H}_\mathsf{opt}\), and additionally untied optimized weights over all iterations and edges. It is important to note that to obtain the untied weights, an extra training step with untied weights is required to obtain \(\mathcal{W}_\mathsf{opt}\).
   \end{itemize}
 All three decoders require the same number of \ac{CN} evaluations as they operate on the same set of parity-check matrices \(\mathcal{H}_\mathsf{opt}\). However, they differ in the required memory. \ac{PBNBP} decoder \(\mathcal{D}_3\) needs to store one weight per edge, whereas \ac{PBNBP} decoder  \(\mathcal{D}_2\) does not need to store any weights. \ac{PBNBP} decoder \(\mathcal{D}_1\) needs to store one weight per channel message and per edge emanating from a \ac{VN} but only one weight per \ac{CN} and hence is of lower complexity compared to  \ac{PBNBP} \(\mathcal{D}_3\) but is more complex than \ac{PBNBP} \(\mathcal{D}_2\).

 \emph{Remark:} We note that in terms of performance, pruning a \ac{CN} is equivalent to setting the weights of all outgoing edges from a \ac{CN} to zero. Hence, an \ac{NBP} decoder could in theory be trained to yield exactly the same performance as a \ac{PBNBP} decoder. However, the \ac{NBP} decoder would have a higher complexity than the \ac{PBNBP} decoder as the number of \ac{CN} evaluations would not be reduced.  Furthermore, empirically, we do not observe that weights converge to zero for \ac{NBP}. In this light, pruning allows for a reduction in complexity as well as facilitates training. This is in line with the results in \cite{Frankle2019}, where the authors observed that by iteratively pruning and retraining a vanilla neural network, the network achieves higher accuracy and converges faster than the original, unpruned network or a network pruned using a single iteration.

\subsection{Pruning-Based Neural Offset Min-Sum Decoder}

We extend the pruning concept to  \ac{NOMS} decoding. The same procedure as the one described in the previous section can be applied to find the optimal set of parity-check matrices and weights for \ac{PBNOMS}. However, since both \ac{NBP} and \ac{NOMS} are iterative algorithms differing only in the \ac{CN} update, we hypothesize that the same set of parity-check matrices optimized for \ac{NBP} performs well for \ac{NOMS}. This is confirmed empirically by comparing the performance of a \ac{PBNOMS} decoder with parity-check matrices optimized using \ac{NBP} and a \ac{PBNOMS} decoder with parity-check matrices optimized using \ac{NOMS}.

%%%%%%%%%%%%%%%%%%%%%%%%%%%%%%%%%%%%%%%%%%%%%%%%%%%%%%%%%%%%%%%%%%%%%%%%%%%%%%%
%                                                                             %
%                                QUANTIZIATION                                %
%                                                                             %
%%%%%%%%%%%%%%%%%%%%%%%%%%%%%%%%%%%%%%%%%%%%%%%%%%%%%%%%%%%%%%%%%%%%%%%%%%%%%%%
\section{Quantization of the PB-NOMS Decoder}
\label{sec:quantization}

We consider quantization of the channel output, the messages between the layers, and the weights and offsets of the \ac{PBNOMS} decoder.
In particular, we consider a joint optimization of  the quantization, the weights, and the offsets. To this end, we first define a symmetric mid-tread quantizer as a piecewise constant function \(Q(x)\) with \(2^{n_q}-1\) quantization levels \(\mathcal{Q} = \left\{q_0=0, \pm q_1, \ldots,\pm q_{2^{(n_q-1)}-1}\right\}\) and thresholds \(\mathcal{T} = \left\{t_1, \ldots, t_{2^{(n_q-1)}-1}\right\}\),

\begin{align*}
  Q(x) = \left\{\begin{array}{cr@{}C@{}C@{}C@{}l}
                    0 &  &  & |x| & < & t_1,\\
                    \sign{x} q_1 & t_1 & \leq & |x| & <&  t_2,\\
                    \vdots       &      &     &  \vdots   &   &\\
                    \sign{x} q_i & t_{i} & \leq & |x| & < & t_{i+1},\\
                    \vdots       &      &     &   \vdots  &   & \\
                    \sign{x} q_{2^{(n_q-1)}-1} & t_{2^{(n_q-1)}-1} & \leq & |x| & . &
                  \end{array}\right.
\end{align*}
The gradients of \(Q(x)\) with respect to \(x\) are zero except at the thresholds. During training in the back-propagation phase, these zero-gradients would cause most gradients in the network to be zero and hence prohibit the training to converge to a meaningful solution.
Using the gradient of a surrogate function, referred to as \ac{STE}, overcomes this issue \cite{Yin2019}. In this work, we simply pass through the gradients with respect to \(x\), i.e., \(\partial Q(x) / \partial x = 1\). By letting \(t_i = (q_{i}+q_{i+1})/2\), the gradients with respect to the quantization levels are
\begin{align*}
  \frac{\partial Q(x)}{\partial q_i} &= \left\{\begin{matrix}
                                                \sign{x} & t_{i} \leq |x| < t_{i+1},\\
                                                0 & \text{else}
                                              \end{matrix}\right.
\end{align*}
and
\begin{align*}
  \frac{\partial Q(x)}{\partial q_{2^{n_q-1}-1}} &= \left\{\begin{matrix}
                                                \sign{x} & t_{2^{(n_q-1)}-1} \leq |x|,\\
                                                0 & \text{else}
                                              \end{matrix}\right.
\end{align*}
 for \(i \in \{1,\ldots 2^{n_q-1}-2\}\).

Denote now as \(Q_\mathsf{ch}^{(\ell)}\), \(Q_{\mathsf{c}\rightarrow\mathsf{v}}^{(\ell)}\), \(Q_{\mathsf{v}\rightarrow\mathsf{c}}^{(\ell)}\), \(Q_{\mathsf{w}_{\mathsf{v}\rightarrow\mathsf{c}}}^{(\ell)}\), and \(Q_\mathsf{\beta}^{(\ell)}\) the quantizers in the \(\ell\)-th iteration for the channel messages, the \ac{CN}-to-\ac{VN} messages, the \ac{VN}-to-\ac{CN} messages, the \ac{VN}-to-\ac{CN} weights, and the offsets corresponding to the CN update (see
 \eqref{eq:cn_update_noms}), respectively.

\begin{figure}
  \centering
  \includegraphics{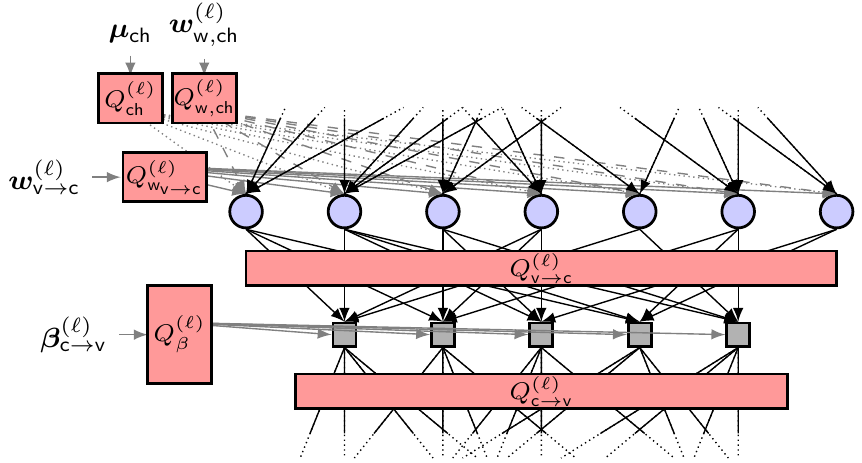}
  \caption{Block diagram of the \(\ell\)-th \ac{VN} and \ac{CN} layer of the unrolled \mbox{(PB-)\ac{NOMS}} decoder with quantization. The blue circles denote \acp{VN}, the gray squares \acp{CN}, \(\bm{\mu}_\mathsf{ch}\) the vector containing all \(n\) channel messages, \(\bm{w}^{(\ell)}_{\mathsf{v}\rightarrow\mathsf{c}}\) and \(\bm{w}^{(\ell)}_{\mathsf{ch}}\) the vectors containing the respective weights, and \(\bm{\beta}^{(\ell)}_{\mathsf{c}\rightarrow\mathsf{v}}\) the vector containing the offsets.}
  \label{fig:quantizer_network}
\end{figure}

In \internalFig{fig:quantizer_network}, we show a block diagram of the \(\ell\)-th \ac{VN} and \ac{VN} layer of a (PB-)\ac{NOMS} decoder. While the forward and backward pass of the training of the quantized \ac{PBNOMS} decoder take into account the quantization, the update of trainable weights during training is performed using floating point precision. To emphasize this, we depict the weights, quantized by its associated quantizer, as an input to the respective node.
Further, while the quantizers are untied over the layers, i.e., each layer has its own quantization levels and thresholds, we tie the number of bits over the layers. We denote the number of bits for quantizing the channel messages as \(q_\mathsf{ch}\), the number of bits for messages passed between nodes as \(q_\mathsf{m}\), and the number of bits for all the weights and offsets as \(q_\mathsf{w}\).

\begin{figure}
    \centering
    \includegraphics{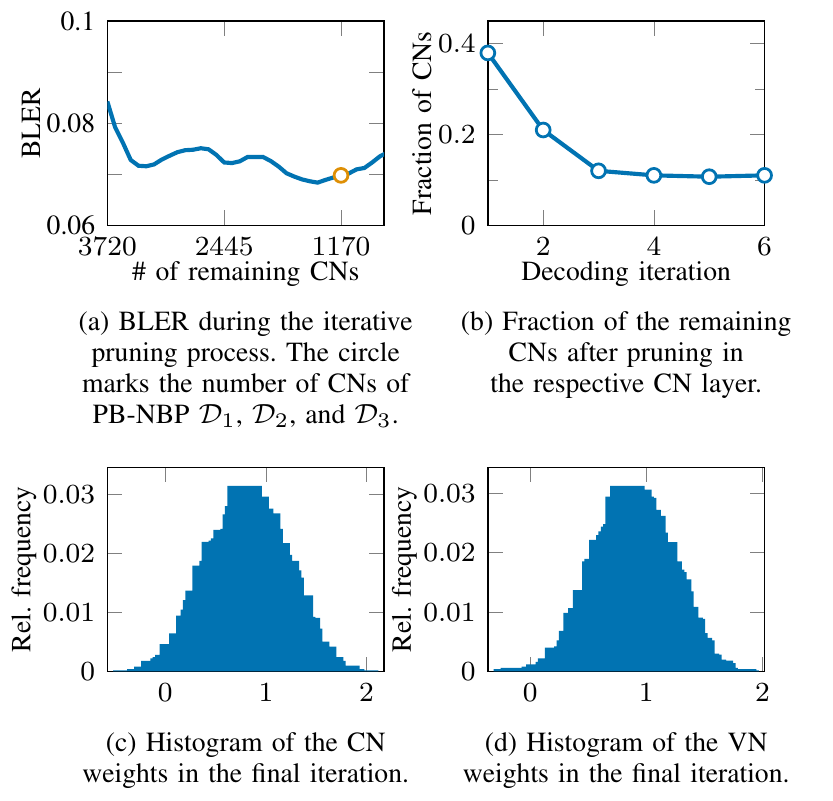}
    \caption{Results from the training process for the RM\((2,5)\) code.}
    \label{fig:rm_2_5_cn_fraction}
\end{figure}

%%%%%%%%%%%%%%%%%%%%%%%%%%%%%%%%%%%%%%%%%%%%%%%%%%%%%%%%%%%%%%%%%%%%%%%%%%%%%%%
%                                                                             %
%                              NUMERICAL RESULTS                              %
%                                                                             %
%%%%%%%%%%%%%%%%%%%%%%%%%%%%%%%%%%%%%%%%%%%%%%%%%%%%%%%%%%%%%%%%%%%%%%%%%%%%%%%
\section{Numerical Results}
\label{sec:numerical}
We numerically evaluate the performance of the proposed \ac{PBNBP} and \ac{PBNOMS} decoders for \ac{RM} codes, a short \ac{LDPC} code, and a polar code. As a benchmark we consider \ac{ML} decoding and we compare the performance of the proposed decoders  to \ac{NBP} \cite{Nachmani2016}, \ac{NOMS} \cite{Lugosch2017}, and \ac{MBBP} decoding \cite{Hehn2010} (referred to as MBBP-NX-S in \cite{Hehn2010}). The hyperparameters are provided in the appendix and the source code is available online \cite{Buchberger2020}.

%-----------------------------------------------------------------------------%
%                                  RM(2,5)                                    %
%-----------------------------------------------------------------------------%
\subsection{\Acl{RM} Code RM\((2,5)\)}

For the  RM\((2,5)\) code of length \(n=32\) and dimension \(k=16\), we consider the overcomplete parity-check matrix \(\bm{H}_\mathsf{oc}\) containing all \(620\) parity-check equations of minimum weight and apply the training procedure described in \internalLinkSection{sec:training} to it. We fix the number of iterations to six. Hence,  without pruning \(620\) \acp{CN} need to  be evaluated per \ac{CN} layer, which leads to a total of \(3720\) \acp{CN} that need to be evaluated. Note that since \(\bm{H}_\mathsf{oc}\) has a regular \ac{CN} degree and pruning does not affect the \ac{CN} degree, we can neglect the average \ac{CN} degree in \internalEq{eq:complexity} for the complexity discussion.
The optimization is stopped when the loss starts to increase, which corresponds to keeping \(\SI{31}{\percent}\) of the \acp{CN} of the unrolled graph. This is shown in Fig. \ref{fig:rm_2_5_cn_fraction}(a), where the \ac{BLER} during the iterative pruning process is shown as a function of the number of remaining \acp{CN}. The marker highlights where the pruning is stopped. Note that this does not correspond to the minimum of the \ac{BLER} as training is only stopped after the loss started to increase (see also \internalLinkSection{sec:training}). To investigate the behavior of pruning, we are interested in how many \acp{CN} are pruned in each \ac{CN} layer. In Fig. \ref{fig:rm_2_5_cn_fraction}(b), we plot the distribution of  surviving \acp{CN} after pruning  across \ac{CN} layers. We observe that in the first \ac{CN} layer, about \(\SI{40}{\percent}\) of all surviving \acp{CN} are used for decoding. In later \ac{CN} layers, the number of \acp{CN} decreases significantly. This observation furthermore justifies the use of a low number of iterations. In Fig.~\ref{fig:rm_2_5_cn_fraction}(c) and \ref{fig:rm_2_5_cn_fraction}(d), we plot the histograms of the \ac{CN} and \ac{VN} weights in the final iteration.

\begin{figure}
    \centering
    \includegraphics{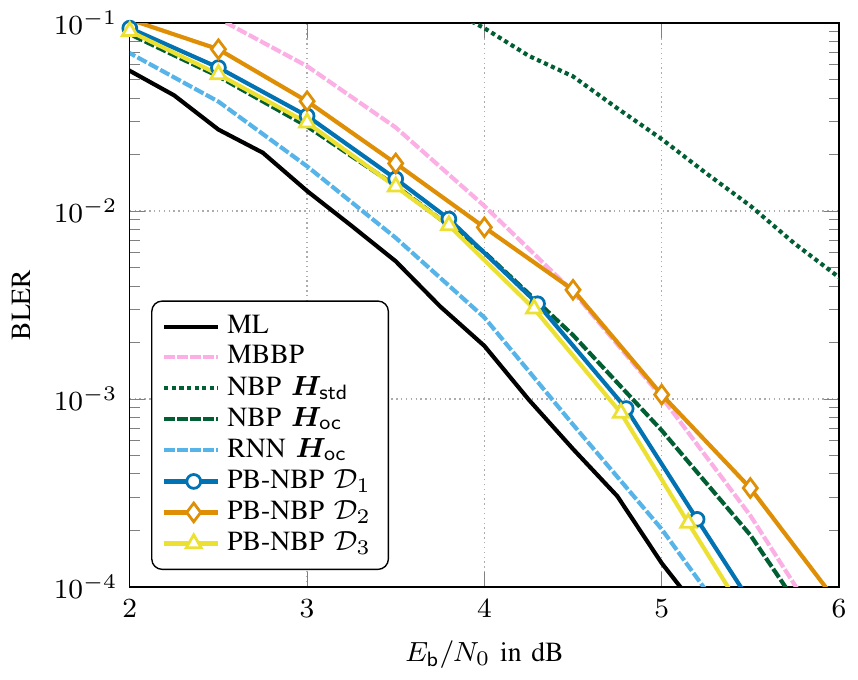}
    \caption{BLER results for the RM\((2, 5)\) code with (PB-)NBP decoding.}
    \label{fig:bler_rm_2_5}
\end{figure}

In \internalFig{fig:bler_rm_2_5}, we plot the \ac{BLER} as a function of \(\EbNoLine\).
The \ac{PBNBP} decoder \(\mathcal{D}_1\) performs  \(\SI{0.38}{\decibel}\) away from the \ac{ML} decoder at a \ac{BLER} of \(10^{-4}\). Removing the weights  (\ac{PBNBP} \(\mathcal{D}_2\)), results in a penalty of \(\SI{0.48}{\decibel}\).  Untying the weights in the \acp{CN} (\ac{PBNBP} \(\mathcal{D}_3\)) results in an additional gain of \(\SI{0.047}{\decibel}\) with respect to \ac{PBNBP} \(\mathcal{D}_1\).
\begin{table}
  \centering
  \caption[Complexity \(\RM(2,5)\) code]{Complexity of the \(\RM(2,5)\) code. In parentheses the fraction of the number of \acp{CN} and weights compared to \ac{NBP} with \(\bm{H}_\mathsf{oc}\) (denoted by \("1.0"\)).}
  \label{tab:complexity_rm_2_5}
  \begin{tabularx}{\tablescaling\columnwidth}{Xrlrl}
    \toprule
     &  \multicolumn{2}{c}{\# of \acp{CN}} & \multicolumn{2}{c}{\# of weights and offsets}\\
    \midrule
    \ac{NBP} \(\bm{H}_\mathsf{oc}\)  & \(3720\) &\((1.0)\) & \(64704\) & \((1.0)\)\\
    RNN \(\bm{H}_\mathsf{oc}\)  & \(3720\) &\( (1.0)\) & \(10144\) & \((0.157)\)\\
    \ac{NBP} \(\bm{H}_\mathsf{std}\)  & \(96\) & \( (0.026)\) & \(328\)&\( (0.005)\)\\
     \ac{MBBP} RM\((2, 5)\) & \(1440\) & \((0.387)\) & \(0\) & \((0.0)\)\\
     \midrule
    \ac{PBNBP} \(\mathcal{D}_1\) & \(1170\) & \((0.315)\) & \(10754\) & \((0.166)\)\\
    \ac{PBNBP} \(\mathcal{D}_2\)  & \(1170\) & \((0.315)\) & \(0\) & \( (0.0)\)\\
    \ac{PBNBP} \(\mathcal{D}_3\)  & \(1170\) & \((0.315)\) & \(18944\)&\((0.293)\)\\
    \ac{PBNBP} Random   & \(1170\) & \((0.315)\) & \(10754\) & \((0.166)\)\\
    \ac{PBNOMS} & \(1170\)&\( (0.315)\) & \(9360\)&\((0.145)\)\\
    \bottomrule
  \end{tabularx}
\end{table}

Both \ac{PBNBP} \(\mathcal{D}_1\) and \(\mathcal{D}_3\) outperform \ac{NBP}  with \(\bm{H}_\mathsf{oc}\) containing the \(620\) parity-check equations of minimum weight, as well as  \ac{MBBP} \cite{Hehn2010} with \(15\) randomly chosen parity-check matrices. Furthermore, the proposed \ac{PBNBP} decoders are less complex than \ac{NBP} with \(\bm{H}_\mathsf{oc}\) and \ac{MBBP}, requiring \(\SI{68}{\percent}\) and \(\SI{7}{\percent}\) less \ac{CN} evaluations than \ac{NBP} and \ac{MBBP}, respectively.
 \ac{PBNBP} \(\mathcal{D}_2\) performs slightly worse than \ac{NBP}, but entails the lowest complexity as no weights need to be stored. As a further comparison, we also plot the performance of a \ac{RNN}-based decoder \cite{Nachmani2018} using \(\bm{H}_\mathsf{oc}\). The \ac{RNN}-based decoder slightly outperforms \ac{PBNBP} \(\mathcal{D}_1\) and \ac{PBNBP} \(\mathcal{D}_2\) at the cost of increased complexity by a factor of three. \ac{NBP} with a standard parity-check matrix with no redundant rows (i.e., containing \(16\) parity-check equations) is clearly not competitive.
The \ac{PBNBP} decoders also require significantly less weights than the \ac{NBP} decoder with \(\bm{H}_\mathsf{oc}\).
The decoding complexity of the decoders in  \internalFig{fig:bler_rm_2_5} is reported in \internalTab{tab:complexity_rm_2_5}. In parentheses, we display the complexity normalized by that of \ac{NBP} with \(\bm{H}_\mathsf{oc}\).

\begin{figure}
    \centering
    \includegraphics{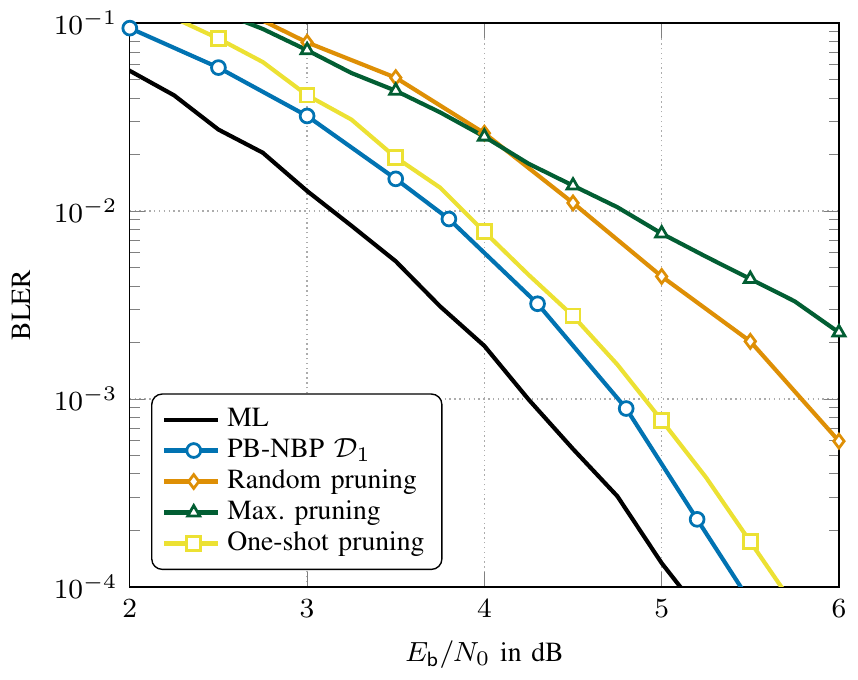}
    \caption{Comparison of different pruning strategies for the RM\((2, 5)\) code.}
    \label{fig:pruning_comparison_rm_2_5}
\end{figure}
To assess the effectiveness of our pruning strategy, we also consider the scenarios where we randomly prune \acp{CN} (random pruning), where we prune the \ac{CN} associated with the largest weight (referred to as maximum pruning), and  all \acp{CN} in a single step (one-shot pruning). As it can be observed in \internalFig{fig:pruning_comparison_rm_2_5}, the performance of random and maximum pruning is clearly not competitive. One-shot pruning exhibits a loss of about \(\SI{0.1}{\decibel}\) over our proposed pruning method. While one-shot pruning may speed up the training process, it is important to note that training is done offline and the final decoders have the same complexity. Furthermore, one-shot pruning requires that the final number of \acp{CN} is known \emph{a priori}.

\begin{figure}
    \centering
    \includegraphics{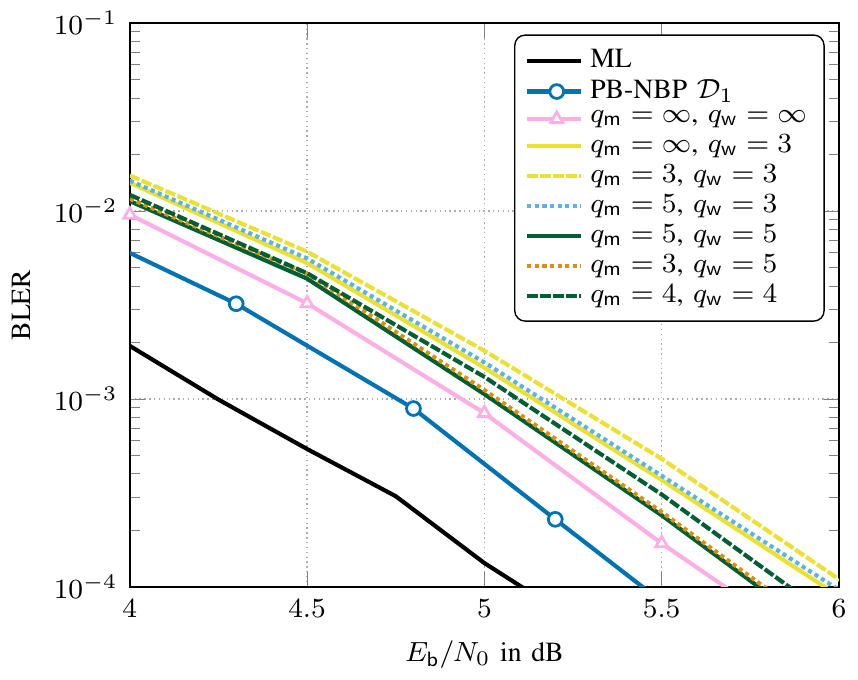}
    \caption{BLER results for the RM\((2, 5)\) code with \ac{PBNOMS} decoding and quantization.}
    \label{fig:bler_rm_2_5_noms_quant}
\end{figure}

The \ac{BLER} of the \ac{PBNOMS} decoder with \(q_\mathsf{m}\) bits for the messages and \(q_\mathsf{w}\) bits for the weights is depicted in \internalFig{fig:bler_rm_2_5_noms_quant}, where \(\infty\) bits denotes full precision floating point. We further set \(q_\mathsf{ch} = q_\mathsf{m}\).
Floating point \ac{PBNOMS} decoding suffers from a \(\SI{0.25}{\decibel}\) degradation over \ac{PBNBP} decoding. Further quantizing the \ac{PBNOMS} decoder increases this gap. We observe that while for \(q_\mathsf{w}=5\), \(3\)~bits for the channel output and internal messages appears to be sufficient, in the case of \(q_\mathsf{w}=3\) using \(3\)~bits for the channel output and internal messages  leads to a significant performance degradation with respect to floating point messages.
The \ac{PBNOMS} and the \ac{PBNBP} decoders require the same number of \ac{CN} evaluations and weights.
However, the \ac{PBNOMS} decoder is less complex, due to its simplified \ac{CN} update.

%-----------------------------------------------------------------------------%
%                                  RM(3,7)                                    %
%-----------------------------------------------------------------------------%
\subsection{\Acl{RM} Code RM\((3,7)\)}
For the RM\((3,7)\) code of code length \(n=128\) and dimension \(k=64\), we select as the overcomplete parity-check matrix \(\bm{H}_\mathsf{oc}\) the matrix containing \(70000\) out of \(94488\) randomly selected parity-check equations of minimum weight. Since in the initial training phase \acp{CN} are removed in an almost random fashion, choosing a large random subset of all minimum-weight parity-check equations to initialize the training does not harm the optimization.
To investigate the effect of the size of the random subsets of minimum-weight parity-check equations, we also consider the case where  only a small, random subset, containing \(9448\)  of all parity-check equations of minimum weight are selected for \(\bm{H}_\mathsf{oc}\) and denote this decoder as \ac{PBNBP} \(\mathcal{\tilde{D}}_1\).
Again, we fix the number of iterations to six.
\begin{figure}
    \centering
    \includegraphics{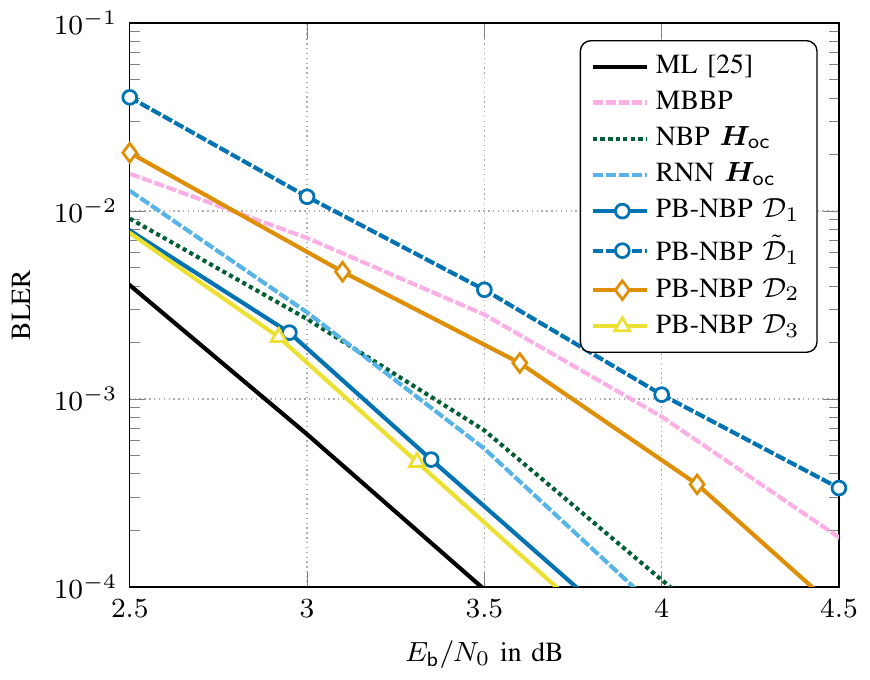}
    \caption{BLER results for the RM\((3, 7)\) code with (PB-)\ac{NBP} decoding.}
    \label{fig:bler_rm_3_7}
\end{figure}
\begin{table}
  \centering
  \caption[Complexity \(\RM(3,7)\) code]{Complexity of the \(\RM(3,7)\) code. In parentheses the fraction of the number of \acp{CN} and weights compared to \ac{NBP} with \(\bm{H}_\mathsf{oc}\) (denoted by \("1.0"\)).}
  \label{tab:complexity_rm_3_7}
  \begin{tabularx}{\tablescaling\columnwidth}{Xrlrl}
    \toprule
     & \multicolumn{2}{c}{\# of \acp{CN}} & \multicolumn{2}{c}{\# of weights and offsets}\\
    \midrule
    \ac{NBP} \(\bm{H}_\mathsf{oc}\) & \(566928\) & \((1.0)\) & \(19654400\) & \((1.0)\)\\
    RNN \(\bm{H}_\mathsf{oc}\) & \(566928\) & \((1.0)\) &  \(3023744\)& \((0.154)\)\\
    \ac{MBBP} & \(23440\) & \((0.0413)\) & \(0\) & \((0.0)\)\\
    \midrule
    \ac{PBNBP} \(\mathcal{D}_1\)  & \(19842\) & \((0.0349)\) & \(2342252\) & \((0.119)\)\\
    \ac{PBNBP} \(\mathcal{D}_2\)  & \(19842\) & \((0.0349)\) & \(0\) & \((0.0)\)\\
    \ac{PBNBP} \(\mathcal{D}_3\)  & \(19842\) & \((0.0349)\) & \(4128032\) & \((0.21)\)\\
    \ac{PBNBP} \(\mathcal{\tilde{D}}_1\)  & \(19842\) & \((0.0349)\) & \(2342252\) & \((0.119)\)\\
   \ac{PBNOMS}  & \(19842\) & \((0.0349)\) & \(1904832\) & \((0.097)\)\\
    \bottomrule
  \end{tabularx}
\end{table}
In \internalFig{fig:bler_rm_3_7}, we plot the \ac{BLER} for the \ac{PBNBP} decoders \(\mathcal{D}_1\), \(\mathcal{\tilde{D}}_1\), \(\mathcal{D}_2\), and \(\mathcal{D}_3\) and compare the performance to that of \ac{NBP} and \ac{MBBP}. Decoder \ac{PBNBP} \(\mathcal{D}_1\) performs  \(\SI{0.27}{\decibel}\) from the \ac{ML} decoder and improves upon \ac{NBP} with \(\bm{H}_\mathsf{oc}\) by \(\SI{0.28}{\decibel}\). Removing the weights results in a degradation of
 \(\SI{0.47}{\decibel}\) for decoder \ac{PBNBP} \(\mathcal{D}_2\) with respect to \ac{PBNBP} \(\mathcal{D}_1\). On the other hand, untying the weights results in a gain of \(\SI{0.02}{\decibel}\) for decoder \ac{PBNBP} \(\mathcal{D}_3\).

 Decoders \ac{PBNBP} \(\mathcal{D}_1\), \ac{PBNBP} \(\mathcal{D}_2\), and \ac{PBNBP} \(\mathcal{D}_3\), require only \(\SI{3.49}{\percent}\) of the \acp{CN} and at most \(\SI{21}{\percent}\) of the weights compared to the \ac{NBP} decoder with \(\bm{H}_\mathsf{oc}\) while showing a performance gain of \(\SI{0.28}{\decibel}\) for decoder \ac{PBNBP} \(\mathcal{D}_3\). The \ac{RNN}-based decoder with \(\bm{H}_\mathsf{oc}\) requires a similar number of weights as the \ac{PBNBP} decoder \(\mathcal{D}_3\), but is significantly more complex as it has the same number of \acp{CN} as the \ac{NBP} decoder. However, it offers a worse performance than \ac{PBNBP} \(\mathcal{D}_1\) and \(\mathcal{D}_3\). As for the \ac{RM}\((2,5)\) code, \ac{NBP} using the non-redundant parity-check matrix with \(64\) \acp{CN} is not competitive (curve omitted for better readability). The complexities are reported in \internalTab{tab:complexity_rm_3_7}. As for the RM\((2,5)\) code, we omit the scaling by the average constant \ac{CN} degree.

Decoder \ac{PBNBP} \(\mathcal{\tilde{D}}_1\) demonstrates the effect of using only a small subset of all parity-check equations of minimum weight as the  overcomplete parity-check matrix \(\bm{H}_\mathsf{oc}\). The decoder is pruned to the same complexity as \ac{PBNBP} \(\mathcal{D}_1\). As randomly selecting a small subset of parity-check equations essentially corresponds to randomly pruning \acp{CN}, we observe the same performance degradation as for the random pruning in the case of the \ac{RM}\((2,5)\) code.

\begin{figure}[t]
     \centering
     \includegraphics{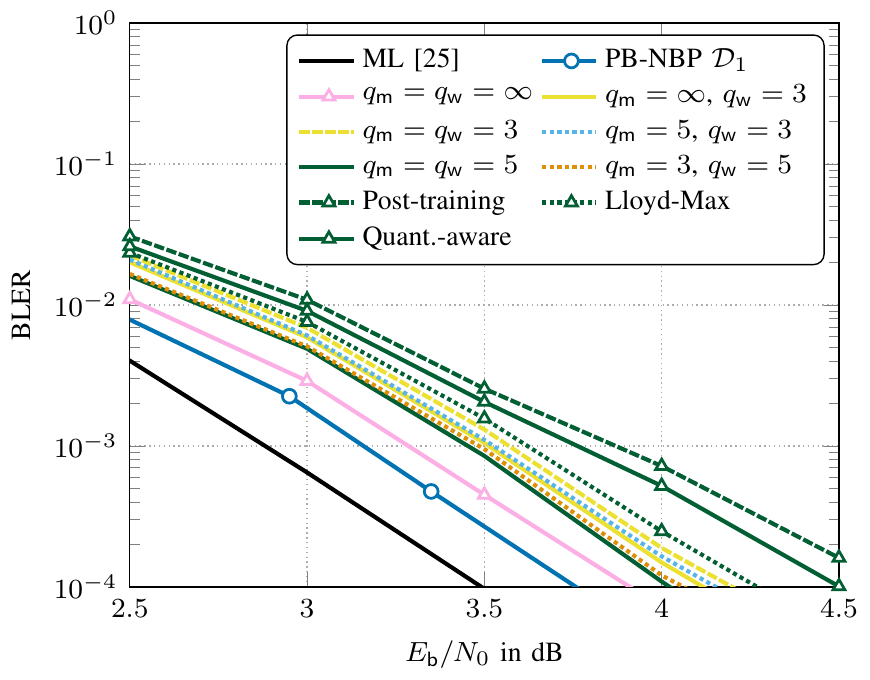}
     \caption{BLER results for the RM\((3, 7)\) code with \ac{PBNOMS} decoding and quantization. For post-training quantization and quantization-aware training, \mbox{\(q_\mathsf{m}=q_\mathsf{ch}=q_\mathsf{w}=3\)} is used.}
     \label{fig:bler_rm_3_7_noms_quant}
 \end{figure}

In \internalFig{fig:bler_rm_3_7_noms_quant}, we report \ac{BLER} results for the quantized and pruned \ac{NOMS} decoders. Once again, we set \(q_\mathsf{ch} = q_\mathsf{m}\). At a \ac{BLER} of \(10^{-4}\), with \(q_\mathsf{m} = q_\mathsf{w} = 5\), we perform \(\SI{0.5}{\decibel}\) from \ac{ML} and with only \(3\)~bits, we perform \(\SI{0.8}{\decibel}\) from \ac{ML}. In the figure we also compare the performance of the joint optimization of  the quantizers, weights, and offsets to the that of two common approaches to quantization in neural networks---\emph{post-training quantization} and \emph{quantization-aware training}. For post-training quantization, the decoder is trained using floating point precision and the quantizer is added after the training is completed. To this end, we use Tensorflow's built-in quantizer, i.e., a uniform quantizer. The clipping range for the messages is set to \(\pm 8\). The quantizer of the weights and offsets is clipped to the range of the weights and offsets in the respective layer. From \internalFig{fig:bler_rm_3_7_noms_quant}, we notice that this way of quantizing is clearly not competitive. One reason for this is that the weights and offsets of the decoder may not be optimal once the quantizers are added, as  quantization distorts both weights and offsets. Furthermore, a different set of weights and offsets may be able to (partially) compensate for the performance degradation due to the quantized messages.
 Incorporating the quantizer into the training, referred to as quantization-aware training, overcomes this. Once again, we use Tensorflow's built-in quantizer for this, clip the messages to \(\pm 8\) and the weights and offsets to the range of the weights and offsets in the respective layer. Even though quantization-aware training improves upon post-training quantization, it is limited by the initial choice of using a uniform quantizer and the clipping range for the messages. For post-training quantization, a quantizer optimized using the Lloyd-Max algorithm \cite{Max1960,Lloyd1982} improves significantly over the uniform quantizers. However, as the weights and offsets are potentially suboptimal, a small degradation to the proposed joint optimization remains.

%-----------------------------------------------------------------------------%
%                                    LDPC                                     %
%-----------------------------------------------------------------------------%
\subsection{Low-Density Parity-Check Code}
We consider the CCSDS \ac{LDPC} code of length \(128\) and rate \(0.5\) as defined in \cite{ccsds}. It has \ac{CN} degree \(8\) and half the \acp{VN} have degree \(3\) and half have degree \(5\). The code has minimum Hamming distance \(14\).
We consider conventional \ac{BP} decoding with \(25\) iterations, corresponding to \(1600\) \ac{CN} updates.

We let the overcomplete parity-check matrix \(\bm{H}_\mathsf{oc}\) contain \(10000\) randomly chosen parity-check equations of Hamming weight up to twenty. This causes \acp{CN} of different degrees and hence \(\bm{H}_\mathsf{oc}\) has irregular \ac{CN} degree. We then design the \ac{PBNBP} decoder with six iterations and the same number of \ac{CN} evaluations as for conventional \ac{BP} decoding with \(25\) iterations. Note that as the parity-check matrices of the \ac{BP} and \ac{PBNBP} decoders are of irregular \ac{CN} degree,  the complexity is given by \internalEq{eq:complexity}.
\begin{table}
  \centering
  \caption[Complexity CCSDS LDPC code]{Complexity of the CCSDS \ac{LDPC} code. In parentheses the fraction of the number of \acp{CN} compared to conventional \ac{BP} with \(100\) decoding iterations (denoted by \("1.0"\)).}
  \label{tab:complexity_ccsds}
  \begin{tabularx}{\tablescaling\columnwidth}{Xrlr}
    \toprule
    & \multicolumn{2}{c}{Complexity \internalEq{eq:complexity}} & \multicolumn{1}{c}{\# of weights and offsets}\\
    \midrule
    BP,  100 iterations  & \(51200\) & \((1.0)\) & \(0\)\\
    BP,  25 iterations  & \(12800\) & \((0.25)\) & \(0\)\\
    \ac{NBP} \(\bm{H}_\mathsf{std}\) & \(768\) & \((0.015)\) & \(13696\)\\
    \midrule
    \ac{PBNBP} \(\mathcal{D}_1\) & \(25920\) & \((0.506)\) & \(28416\)\\
    \ac{PBNOMS}  & \(25920\) & \((0.506)\) & \(25920\)\\
    \bottomrule
  \end{tabularx}
\end{table}
\begin{figure}
    \centering
    \includegraphics{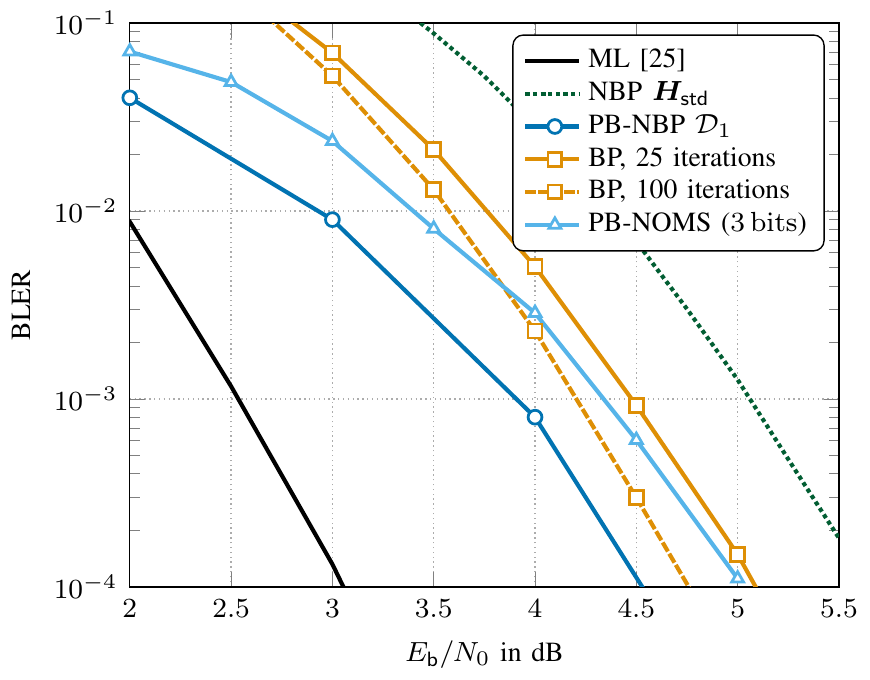}
    \caption{\Ac{BLER} results for the LDPC code.}
    \label{fig:bler_ldpc}
\end{figure}

The \ac{BLER} results are given in \internalFig{fig:bler_ldpc}. \Ac{PBNBP} decoder \(\mathcal{D}_1\) outperforms conventional \ac{BP} with \(25\) iterations by approximately \(\SI{0.6}{\decibel}\).
Allowing \(100\) iterations for conventional \ac{BP} reduces the gain to \(\SI{0.2}{\decibel}\). However, conventional \ac{BP} with \(100\) iterations requires approximately twice the complexity of decoder \ac{PBNBP} \(\mathcal{D}_1\). \ac{NBP} with \(64\) \acp{CN} is again not competitive. The complexity of the different decoders is reported in \internalTab{tab:complexity_ccsds}. We remark that the decoding complexity of \ac{BP} decoding does not take into account potential early stopping of the decoder. However, early stopping may also be used in the proposed \ac{PBNBP} decoders.
% In   \internalFig{fig:bler_ldpc_quant}, we give results for different quantizations.
The effects of quantization are similar to the ones observed for the Reed-Muller codes. Quantizing messages, channel output, weights, and offsets with \(3\) bits causes a degradation of \(\SI{0.3}{\decibel}\) over \ac{PBNBP} \(\mathcal{D}_1\), which corresponds to a performance of \(\SI{1.8}{\decibel}\) from \ac{ML} decoding.

%-----------------------------------------------------------------------------%
%                                    Polar                                     %
%-----------------------------------------------------------------------------%
\subsection{Polar Code}
\begin{table}[t]
  \centering
  \caption[Complexity polar code]{Complexity of the polar code. In parentheses the fraction of the number of \acp{CN} and weights compared to \ac{NBP} with \(\bm{H}_\mathsf{oc}\) (denoted by \("1.0"\)).}
  \label{tab:complexity_polar}
  \begin{tabularx}{\tablescaling\columnwidth}{Xrlrl}
    \toprule
     & \multicolumn{2}{c}{\# of \acp{CN}} & \multicolumn{2}{c}{\# of weights and offsets}\\
    \midrule
    \ac{NBP} \(\bm{H}_\mathsf{oc}\) & \(589824\) & \((1.0)\) & \(20448128\) & \((1.0)\)\\
    \ac{NBP} \(\bm{H}_\mathsf{std}\) & \(384\) & \((0.00065)\) & \(14208\) & \((0.00069)\)\\
    \midrule
    \ac{PBNBP} \(\mathcal{D}_1\)  & \(19842\) & \((0.033)\) & \(2342252\) & \((0.115)\)\\
    \ac{PBNBP} \(\mathcal{D}_2\)  & \(19842\) & \((0.033)\) & \(0\) & \((0.0)\)\\
   \ac{PBNOMS}  & \(19842\) & \((0.033)\) & \(1904832\) & \((0.064)\)\\
    \bottomrule
  \end{tabularx}
\end{table}
We finally consider a polar code of length \(n=128\) and rate \(0.5\) defined in \cite{channelcodes}. It has minimum Hamming distance  \(8\) and its dual code has minimum Hamming distance \(16\). Following \cite{Bardet2016}, we find  all \(98304\) codewords of the dual code of minimum weight. As for the RM\((3,7)\) code, \(\bm{H}_\mathsf{oc}\) used for the optimization process contains \(70000\) randomly-selected minimum-weight parity-check equations to reduce the complexity of the training. As a target complexity, we choose the same complexity as for the RM\((3,7)\) code. The final complexities are reported in \internalTab{tab:complexity_polar}. Once again, we omit the scaling by the average constant \ac{CN} degree.

\begin{figure}[t]
  \centering
  \includegraphics{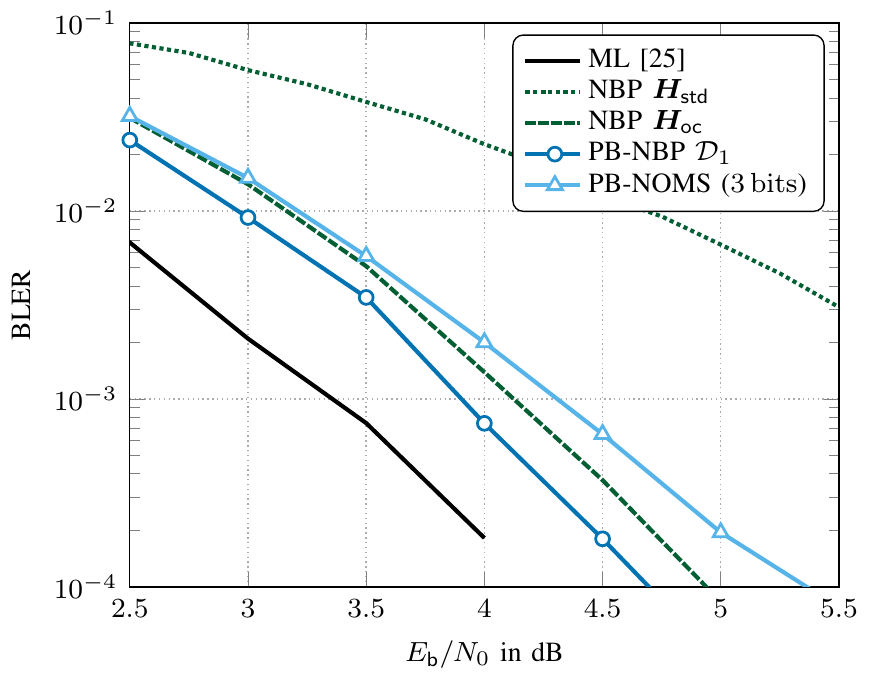}
  \caption{\Ac{BLER} results for the polar code.}
  \label{fig:bler_polar}
\end{figure}

In \internalFig{fig:bler_polar}, we plot the \ac{BLER} as a function of \(\EbNoLine\). We observe a similar behavior to that of the \ac{RM} codes. The \ac{PBNBP} decoder \(\mathcal{D}_1\) performs \(\SI{0.5}{\decibel}\) from \ac{ML} and outperforms \ac{NBP} with \(\bm{H}_\mathsf{oc}\) while only requiring \(\SI{3.3}{\percent}\) of its \acp{CN} evaluations and \(\SI{20.2}{\percent}\) of the weights. Switching to a quantized \ac{PBNOMS} decoder with \(q_\mathsf{w}=q_\mathsf{ch}=q_\mathsf{m}=3\) causes a performance loss of \(\SI{0.65}{\decibel}\). As previously, \ac{NBP} with \(\bm{H}_\mathsf{std}\) offers the lowest complexity, but is clearly not competitive.

%%%%%%%%%%%%%%%%%%%%%%%%%%%%%%%%%%%%%%%%%%%%%%%%%%%%%%%%%%%%%%%%%%%%%%%%%%%%%%%
%                                                                             %
%                                  CONCLUSION                                 %
%                                                                             %
%%%%%%%%%%%%%%%%%%%%%%%%%%%%%%%%%%%%%%%%%%%%%%%%%%%%%%%%%%%%%%%%%%%%%%%%%%%%%%%
\section{Conclusion}
\label{sec:conclusion}
\acreset{PBNBP,PBNOMS}

We proposed a novel \ac{PBNBP} decoder for short linear block codes. The proposed decoder is based on pruning a large overcomplete parity-check matrix and uses different parity-check equations in each decoding iteration.
For Reed-Muller codes and a polar code, we demonstrated  a performance close to \ac{ML} decoding. For a short, standardized \ac{LDPC} code, we showed that we can outperform conventional \c{BP} decoding at a reduced complexity.
 The proposed \ac{PBNBP} decoders outperform \ac{NBP} with large overcomplete parity-check matrices while providing a lower complexity. In the case of a \((3,7)\) Reed-Muller code, the \ac{PBNBP} decoder outperforms the \ac{NBP} decoder by \(\SI{0.28}{\decibel}\) at a \ac{BLER} of \(10^{-4}\) while only requiring \(\SI{3}{\percent}\) of the check nodes and \(\SI{21}{\percent}\) of the weights. Further, it performs within \(\SI{0.27}{\decibel}\) from \ac{ML} decoding. In all scenarios, our approach outperforms the original \ac{NBP} and \ac{MBBP}.
We also applied the proposed pruning-based decoder to \ac{NOMS} and introduced a quantized \ac{PBNOMS} decoder which allows joint optimization of the weights, offsets, and quantization. With messages, weights, and offsets quantized with \(5\)~bits, \ac{PBNOMS} achieves a performance  \(\SI{0.5}{\decibel}\) and  \(\SI{1.7}{\decibel}\) from \ac{ML} for a \acl{RM} and \ac{LDPC} code, respectively. A polar code with \(3\) bits quantization  performs \(\SI{1}{\decibel}\) away from \ac{ML}.
The proposed approach can readily be applied to other linear block codes such as \ac{BCH} codes, with similar gains over the original \ac{NBP} decoder expected. Furthermore, additional constraints can be introduced in the training process to allow for a more practical decoder.

%%%%%%%%%%%%%%%%%%%%%%%%%%%%%%%%%%%%%%%%%%%%%%%%%%%%%%%%%%%%%%%%%%%%%%%%%%%%%%%
%                                                                             %
%                          APPENDIX HYPERPARAMETERS                           %
%                                                                             %
%%%%%%%%%%%%%%%%%%%%%%%%%%%%%%%%%%%%%%%%%%%%%%%%%%%%%%%%%%%%%%%%%%%%%%%%%%%%%%%
\appendix[Hyperparameters]
% \section{}
\label{app:hp}
The \ac{PBNBP} decoders are trained with a batch size of \(128\) in the case of the Reed-Muller code \(\text{RM}(2,5)\) and the \ac{LDPC} code, and a batch size of \(64\) for the Reed-Muller code \(\text{RM}(3,7)\) and the polar code. As an optimizer, the Adam optimizer with a learning rate of \(0.001\) is employed. Initially, \(\eta=1.0\) and every \(3000\) batches \(\eta\) is decreased by multiplying it by \(0.8\). After each pruning step, \(\eta\) is reset to its initial value. The maximum number of batches per pruning step is  \(10^5\) for the \(\text{RM}(2,5)\) and the \ac{LDPC} code, and \(2\cdot 10^5\) for the \(\text{RM}(3,7)\) and the polar code, but the next pruning step is performed earlier if the average loss over \(100\) batches stops decreasing.

The \ac{PBNOMS} decoders are trained starting from the pruned parity-check matrices. For  the \(\text{RM}(2,5)\) and the \ac{LDPC} code, we use \(1.5\cdot 10^5\) batches and for the \(\text{RM}(3,7)\) and the polar code \(3\cdot 10^5\) batches. As for the \ac{PBNBP} decoders, the learning rate is set to \(0.001\) and every \(3000\) batches \(\eta\) is decreased by multiplying it by \(0.8\).

\balance
%%%%%%%%%%%%%%%%%%%%%%%%%%%%%%%%%%%%%%%%%%%%%%%%%%%%%%%%%%%%%%%%%%%%%%%%%%%%%%%
%                                                                             %
%                                  REFERENCES                                 %
%                                                                             %
%%%%%%%%%%%%%%%%%%%%%%%%%%%%%%%%%%%%%%%%%%%%%%%%%%%%%%%%%%%%%%%%%%%%%%%%%%%%%%%
% \nocite*{}
\bibliographystyle{IEEEtran}
\bibliography{IEEEabrv,conference_abbreviations,literature}

\end{document}